\documentclass[a4paper,12pt]{article}
\usepackage{epsfig,psfrag}
\usepackage{citesort}
\date{\today}
 \normalsize

\newcommand{\bmat}{\left(\begin{array}}
\newcommand{\emat}{\end{array}\right)}
\newcommand{\be}{\begin{equation}}
\newcommand{\ee}{\end{equation}}
\newcommand{\bea}{\begin{eqnarray}}
\newcommand{\eea}{\end{eqnarray}}

\def\lsim{\raise0.3ex\hbox{$\;<$\kern-0.75em\raise-1.1ex\hbox{$\sim\;$}}}
\def\gsim{\raise0.3ex\hbox{$\;>$\kern-0.75em\raise-1.1ex\hbox{$\sim\;$}}}

\def\slash#1{\ooalign{\hfil/\hfil\crcr$#1$}}

\begin{document}
\renewcommand{\thefootnote}{\fnsymbol{footnote}}
\rightline{IPPP/02/70} \rightline{DCPT/02/140}
\vspace{.3cm} 
{\Large
\begin{center}
{\bf On supersymmetric contributions to the CP asymmetry of  
the $B \to \phi K_S$ }
\end{center}}
\vspace{.3cm}

\begin{center}
S. Khalil$^{1,2}$ and E. Kou$^{1}$\\
\vspace{.3cm}
$^1$\emph{IPPP, Physics Department, Durham University, DH1 3LE,
Durham,~~U.~K.}
\\
$^2$ \emph{Ain Shams University, Faculty of Science, Cairo, 11566,
Egypt.}

\end{center}

\vspace{.3cm}
\hrule \vskip 0.3cm
\begin{center}
\small{\bf Abstract}\\[3mm]
\end{center}
We analyse the CP asymmetry of the $B \to \phi K_S$ process 
in general supersymmetric models. 
In the framework of the mass insertion approximation, we derive model 
independent limits for the mixing CP asymmetry. 
We show that chromomagnetic type of operator may play an important role in 
accounting for the deviation of the mixing CP asymmetry between $B \to \phi 
K_S$ and $B \to J/\psi K_S$ processes observed by Belle and BaBar experiments.
A possible correlation between the direct and mixing CP asymmetry 
is also discussed. Finally, we apply our result in minimal supergravity model and 
supersymmetric models with non-universal soft terms. 

\begin{minipage}[h]{14.0cm}
\end{minipage}
\vskip 0.3cm \hrule \vskip 0.5cm
%
\section{{\large \bf Introduction}}

With the advent of experimental data from the $B$ factories, the 
Standard Model (SM) will be subject to a very stringent test, with the 
potential for probing virtual effects from new physics. 
Measurements of the CP asymmetries in various processes are at the center of attentions 
since in the SM, all of them have to be consistently explained by a single parameter, 
the phase in the Cabbibo--Kobayashi--Maskawa mixing matrix 
$\delta_{KM}$ ~\cite{KM}.

The BaBar \cite{babar} and Belle \cite{belle} measurements of time dependent asymmetry 
in $B \to J/\psi K_S$  have provided the first evidence for 
the CP violation in the $B$ system. The world average of these results, 
$S_{J/\psi K_S}=\sin 2\beta (2\phi_1 ) = 0.734 \pm 0.054$, is in a good agreement with the SM 
prediction. Therefore, one may have already concluded that 
the KM mechanism is the dominant source of the CP violation in $B$ system. 
However, in the $B \to J/\psi K_S$ process, 
new physics (NP) effects enter only at the one loop level while 
the SM contributions has dominant tree level contributions. Thus, it is natural that  
NP does not show up clearly in this process. 
In fact,  
in order that a significant supersymmetric contribution appear to 
$S_{J/\psi K_S}$, we need a large 
flavour structure and/or large SUSY CP violating phases, which usually do not exist 
in most of the supersymmetric models such as SUSY models with minimal flavour 
violation or SUSY models with non-minimal flavour and hierarchical Yukawa 
couplings (see Ref. \cite{GK} more in detail).

Unlike the $B \to J/\psi K_S$ process, 
$B \to \phi K_S$ is induced only at the one loop level both in SM and NP. 
Thus, it is tempting to expect that the SUSY contributions to this
decay are more significant \cite{SUSY_BphiK1,SUSY_BphiK2,SUSY_BphiK3,NirICHEP}. Based on the KM mechanism of CP 
violation, both CP asymmetries of $B \to \phi K_S$ and $B \to J/\psi K_S$ 
processes should measure $\sin 2 \beta$ with negligible hadronic uncertainties 
(up to $\mathcal{O} (\lambda^2)$ effects, with $\lambda$ being
the Cabbibo mixing). However, the recent measurements by 
BaBar and Belle collaborations show a $2.7 \sigma$ deviation from the 
observed value of $S_{J/\psi K_S}$ \cite{babar2,belle}. 
The average of these two measurements implies 
\be
S_{\phi K_S} = -0.39 \pm 0.41.
\ee
This difference between $S_{J/\psi K_S}$ and $S_{\phi K_S}$ is considered as 
a hint for NP, in particular for supersymmetry. Several works in this  
respect are in the literature with detail discussion on the possible implications 
of this result \cite{SUSY_BphiK4,SUSY_BphiK5,SUSY_BphiK6,SUSY_BphiK7,SUSY_BphiK8,SUSY_BphiK9,SUSY_BphiK10,SUSY_BphiK11,SUSY_BphiK12,SUSY_BphiK13,SUSY_BphiK14,SUSY_BphiK15}. 

As known, in supersymmetric models there are additional sources 
of flavour structures and  CP violation with a strong correlation 
between them. Therefore, SUSY emerges as the natural candidate to 
solve the problem of the discrepancy between the CP asymmetries
$S_{J/\psi K_S}$ and $S_{\phi K_S}$. However, the unsuccessful searches of 
the electric dipole moment (EDM) of electron, neutron, and mercury atom
impose a stringent constraint on SUSY CP violating phases \cite{phase:edm}.
It was shown that the EDM can be naturally suppressed in SUSY models
with small CP phases \cite{phase:edm} or in SUSY models with flavour 
off--diagonal CP violation \cite{phase:edm,hermitian}. It is worth mentioning 
that the scenario of small CP phases $(\lsim 10^{-2})$ 
in supersymmetric models is still allowed by the present experimental 
results \cite{ABK}. In this class of models, the large flavour mixing is 
crucial to compensate for the smallness of the CP phases.

The aim of this paper is to investigate, in a model independent way, the 
question of whether supersymmetry can significantly modify the CP asymmetry 
in the $B \to \phi K_S$ process. We focus on the gluino contributions to the 
CP asymmetry $S_{\phi K_S}$ for the following two reasons. First, it 
is less constrained by the experimental results on the branching ratio of 
the inclusive transitions $B\to X_s \gamma$ and $B \to X_sl^+l^-$ 
than the chargino contributions ~\cite{GabKhal}. 
Second, it includes the effect of the chromomagnetic operator
which, as we will show, has a huge enhancement in SUSY models~\cite{Kagan,Ciuchini:1996vw}.
We perform this analysis at the NLO accuracy in QCD by using the results
of Ali and Greub \cite{AG}. We also apply our result in minimal supergravity 
model  where the soft SUSY breaking terms are universal
and general SUSY models with non--universal soft terms and Yukawa couplings 
with large mixing. 

The paper is organised as follows. In section 2, we present the CP 
violation master formulae in $B$--system including the SUSY contribution. 
In section 3, we discuss the effective Hamiltonian for $\Delta 
B=1$ transition. Section 4 is devoted to the study of the supersymmetric 
contributions to the mixing and direct CP asymmetry $S_{\phi K_S}$ and 
$C_{\phi K_S}$. We show that the chromomagnetic operator plays a crucial 
role in explaining the observed discrepancy between $S_{\phi K_S}$ and $S_{\psi 
K_S}$. In section 5, we analyse the SUSY contributions to $S_{\phi K_S}$ in 
explicit models. We show that only in SUSY models with 
non--universal soft breaking terms and large Yukawa mixing, one can get 
significant SUSY contributions to $S_{\phi K_S}$. Our conclusions
are presented in section 6.

\section{{\large \bf The CP Violation in $B \to \phi K_S$ Process}}
We start the sections by summarising our convention for the 
CP asymmetry in B system. The time dependent CP asymmetry for $B\to \phi K_S$ can 
be described by ~\cite{BS}: 
\begin{eqnarray}
a_{\phi K_S}(t)&=&\frac{\Gamma (\overline{B}^0(t)\to\phi K_S)-\Gamma 
(B^0(t)\to\phi K_S)} {\Gamma (\overline{B}^0(t)\to\phi K_S)+\Gamma (B^0(t)
\to\phi K_S)} \\
&=&C_{\phi K_S}\cos\Delta M_{B_d}t+S_{\phi K_S}\sin\Delta M_{B_d}t
\end{eqnarray}
where $C_{\phi K_S}$ and $S_{\phi K_S}$ represent 
the direct  and the mixing CP asymmetry, respectively and they are given by  
\begin{equation}
C_{\phi K_S}=\frac{|\overline{\rho}(\phi K_S)|^2-1}{|\overline{\rho}(\phi K_S)|^2+1}, 
\ \ 
S_{\phi K_S}=\frac{2Im \left[\frac{q}{p}~\overline{\rho}(\phi K_S)\right]} 
{|\overline{\rho}(\phi K_S)|^2+1}.  \label{eq:SCdef}
\end{equation}
The parameter $\overline{\rho}(\phi K_S)$ is defined by 
\begin{equation}
\overline{\rho} (\phi K_S)=\frac{\overline{A}(\phi K_S)}{A(\phi K_S)}. 
\end{equation}
where $\overline{A}(\phi K_S)$ and $A(\phi K_S)$ are respectively 
the decay amplitudes of $\overline{B}^0$ and $B^0$  meson which can be  
written in terms of the matrix element of the $\Delta B=1$ transition as
\begin{equation}
\overline{A}(\phi K_S)=\langle \phi K_S| \mathcal{H}^{\mbox{eff}}_{\Delta B=1}|\overline{B}^0\rangle,  \ \ \ 
A(\phi K_S)=\langle \phi K_S| \mathcal{H}^{\mbox{eff}\dagger}_{\Delta B=1}|B^0\rangle. 
\end{equation} 
The mixing parameters $p$ and $q$ are defined by 
$|B_1\rangle =p|B^0\rangle +q|\overline{B}^0\rangle ,    
\ \ |B_2\rangle =p|B^0\rangle -q|\overline{B}^0\rangle$
where $|B_{1(2)}\rangle$ are mass eigenstates of $B$ meson. 
The ratio  $q/p$ can be written 
by using the off-diagonal element of the mass matrix and 
its non-identity  $q/p\neq 1$ is the signature of the CP violation through mixing:   
\begin{equation}
\frac{q}{p}=\sqrt{\frac{M_{12}^*-\frac{i}{2}\Gamma_{12}^*}{M_{12}-\frac{i}{2}\Gamma_{12}}}. 
\end{equation}
The off-diagonal element of the mass matrix is given by the matrix element 
of the $\Delta B=2$ transition as 
\begin{equation}
\langle B^0|\mathcal{H}^{\mbox{eff}\dagger}_{\Delta B=2}|\overline{B}^0\rangle \equiv
M_{12}-\frac{i}{2}\Gamma_{12}. 
\end{equation}
In SM, the major contribution to this matrix element is obtained from the  box diagram   
with $W$-gauge boson and top quark in the loop. As a result, we obtain: 
\begin{equation}
\frac{q}{p}=\frac{V_{tb}^*V_{td}}{V_{tb}V_{td}^*}. 
\end{equation}
where we ignored terms $\mathcal{O}(\Gamma_{12}/M_{12})$. 
Since 
$\overline{\rho}(\phi K_S)=\frac{\overline{A}^{SM}(\phi K_S)}{A^{SM}(\phi K_S)}=\frac{V_{tb}V_{ts}^*}{V_{tb}^*V_{td}}=1$, 
the mixing CP asymmetry in $B \to \phi K_S$ process is found to be 
\begin{equation}S_{\phi K_S}=\sin 2\beta . \end{equation}
Therefore, the mixing CP asymmetry in  $B \to \phi K_S$ is same as 
the one in $B \to J/\psi K_S$ process in SM.  

In supersymmetric theories, there are new 
contributions to the mixing parameters through other box diagrams with 
gluinos and charginos exchanges.  These
contributions to the $\Delta B=2$ transition are often parametrised by ~\cite{para1,para2} 
\begin{equation} 
\sqrt{\frac{M_{12}}{M_{12}^{SM}}}\equiv r_de^{i\theta_d},\end{equation}  where $M_{12}=M_{12}^{SM}+M_{12}^{SUSY}$. In this case, 
the ratio of the mixing parameter $q/p$ can be written as  \begin{equation}
\frac{q}{p}=e^{-2i\theta_d}\frac{V_{tb}^*V_{td}}{V_{tb}V_{td}^*}. \end{equation}
Thus, in the framework of SUSY, the mixing CP asymmetry in $B\to J/\psi K_S$ 
is modified as 
\begin{equation}S_{J/\psi K_S}=\sin (2\beta+2\theta_{d}). \end{equation}

\begin{figure}[t]
\begin{center}\scalebox{0.8}{
\includegraphics[width=20cm,height=10cm,keepaspectratio]{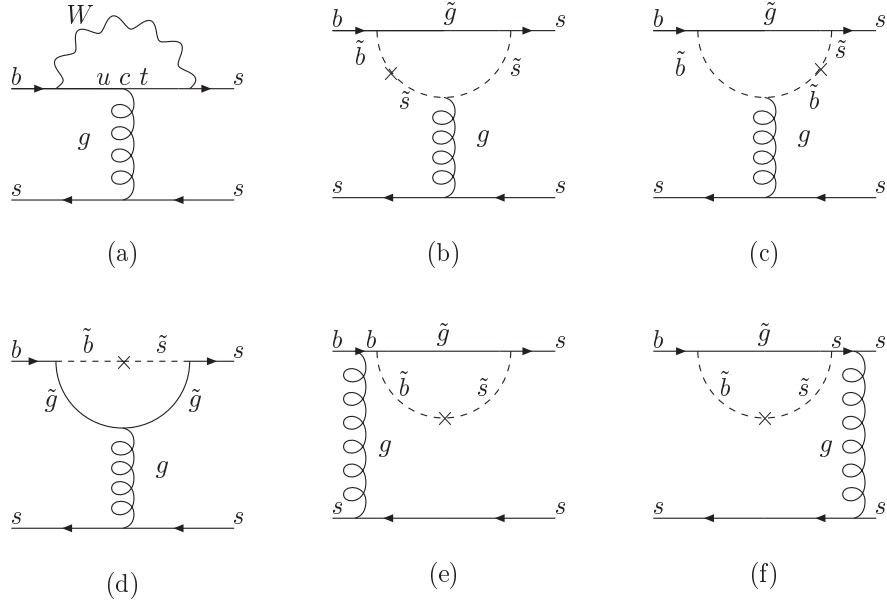}}
\caption{{\small The SM contribution (a) and the gluino--down squark 
contributions (b)--(f) to the
$B \to \phi K_S$ decay.}}
\label{Fig:1}
\end{center}
\end{figure}

In $B\to \phi K_S$ process, we have to additionally consider the SUSY contributions  
to the $\Delta B=1$ transition. 
The supersymmetric contributions to the $\Delta B=1$ transition comes from  
the penguin diagrams with gluinos and charginos in the loop (see Fig.\ref{Fig:1}). 
We can parametrise this effect in the same manner ~\cite{SUSY_BphiK1,para2}: 
\begin{equation}
\frac{A(\phi K_S)}{A^{SM}(\phi K_S)}=S_Ae^{i\theta_A},  
\end{equation}
where $A(\phi K_S)=A^{SM}(\phi K_S)+A^{SUSY}(\phi K_S)$. Therefore,  
we obtain $\overline{\rho} (\phi K_S)=e^{-2i\theta_A}$, hence 
Eq. (\ref{eq:SCdef}) leads to  
\begin{equation}
C_{\phi K_S}=0,  \ \ \  S_{\phi K_S}=
\sin (2\beta +2\theta_d+2\theta_A). 
\end{equation}
However, this parametrisation is true only when we ignore the so-called 
strong phase. 
Since the Belle collaboration  observed nonzero value for $C_{\phi K_S}$ \cite{belle}
we should consider the strong phase in the analysis. In this respect, we reparametrise the 
SM and SUSY amplitudes as 
\begin{eqnarray}
&A^{SM}(\phi K_S)=|A^{\mathrm{SM}}| e^{i\delta_{SM}}, \ \ \ \ 
A^{SUSY}(\phi K_S)=|A^{\mathrm{SUSY}}| e^{i\theta_{SUSY}}e^{i\delta_{SUSY}},  & \\
&\overline{A}^{SM}(\phi K_S)=|\overline{A}^{\mathrm{SM}}| e^{i\delta_{SM}}, \ \ \ \ 
\overline{A}^{SUSY}(\phi K_S)=|\overline{A}^{\mathrm{SUSY}}| e^{-i\theta_{SUSY}}e^{i\delta_{SUSY}},  & 
\end{eqnarray}
where $\delta_{SM(SUSY)}$ is the strong phase (CP conserving) and $\theta_{SUSY}$ is 
the CP violating phase. By using this parametrisation, Eq. (\ref{eq:SCdef}) leads to 
\begin{eqnarray}
S_{\phi K_S} &=& \frac{\sin 2\beta+2\left(\frac{|A^{\mathrm{SUSY}}|}{|A^{\mathrm{SM}}|}\right)\cos
\delta_{12}\sin (\theta_{SUSY}+2\beta )
+\left(\frac{|A^{\mathrm{SUSY}}|}{|A^{\mathrm{SM}}|}\right)^2\sin (2\theta_{SUSY}+2\beta )}
{1+2\left(\frac{|A^{\mathrm{SUSY}}|}{|A^{\mathrm{SM}}|}\right)\cos\delta_{12}\cos\theta_{SUSY}
+\left(\frac{|A^{\mathrm{SUSY}}|}{|A^{\mathrm{SM}}|}\right)^2}, \label{eq:Sfull} \\
C_{\phi K_S} &=& -\frac{2\left(\frac{|A^{\mathrm{SUSY}}|}{|A^{\mathrm{SM}}|}\right)\sin\delta_{12}
\sin\theta_{SUSY}}
{1+2\left(\frac{|A^{\mathrm{SUSY}}|}{|A^{\mathrm{SM}}|}\right)\cos\delta_{12}\cos\theta_{SUSY}
+\left(\frac{|A^{\mathrm{SUSY}}|}{|A^{\mathrm{SM}}|}\right)^2},  \label{eq:Cfull}
\end{eqnarray}
where $\delta_{12}\equiv \delta_{SM}-\delta_{SUSY}$. 
Assuming that the SUSY contribution to the amplitude is smaller than the SM one, 
we can simplify this formula by expanding it in terms of $|A^{\mathrm{SUSY}}|/|A^{\mathrm{SM}}|$
~\cite{Gro}: 
\begin{eqnarray}
S_{\phi K_S}&=& \sin 2\beta 
+2\cos 2\beta\sin\theta_{SUSY}\cos\delta_{12}\frac{|A^{\mathrm{SUSY}}|}{|A^{\mathrm{SM}}|},  \label{eq:Sexp}
\\
C_{\phi K_S}&=& -2\sin \theta_{SUSY}\sin\delta_{12}\frac{|A^{\mathrm{SUSY}}|}{|A^{\mathrm{SM}}|}, \label{eq:Cexp}
\end{eqnarray}
where $\mathcal{O}((|A^{\mathrm{SUSY}}|/|A^{\mathrm{SM}}|)^2)$ is ignored. 
However, as can be seen by comparing these formulae to the measured value of $\sin 2\beta$ 
and the following Belle measurements 
\begin{eqnarray}
S_{\phi K_S}&=&-0.73\pm0.66 , \\ 
C_{\phi K_S}&=&-0.56\pm0.43,  
\end{eqnarray}
a large value of $|A^{\mathrm{SUSY}}|/|A^{\mathrm{SM}}|$ seems to be required. Therefore,  
in our analysis, we use the complete expressions for $S_{\phi K_S}$ and $C_{\phi K_S}$ 
as given in Eqs. (\ref{eq:Sfull}) and  (\ref{eq:Cfull}), respectively. 

Finally, we comment on the branching ratio of the $B^0\to \phi K^0$ 
decay. Since we consider a reasonably large  $|A^{\mathrm{SUSY}}|/|A^{\mathrm{SM}}|$, 
one may wonder that  
the prediction of the branching ratio could be significantly affected. 
The SUSY effect to the branching ratio can be expressed by
\begin{equation}
Br(B^0\to \phi K^0) = Br_{SM}(B^0\to \phi K^0)\left[1+2\cos (\theta_{SUSY}-\delta_{12})\frac{|A_{SUSY}|}{|A_{SM}|}+\left(\frac{|A_{SUSY}|}{|A_{SM}|}\right)^2\right]
\end{equation}
where $Br_{SM}(B^0\to \phi K^0)$ is the standard model prediction of the branching ratio and 
obtained as $(5\sim 8)\times 10^{-6}$.  
The branching ratio is measured by $B$ factories as 
\begin{eqnarray}
Br(B^0\to \phi K^0)&=&(10.0^{\ +1.9\ +0.9}_{\ -1.7\ -1.3})\times 10^{-6}  \ \ \ \mbox{Belle\ \ }~\cite{BrBelle}\\
Br(B^0\to \phi K^0)&=&(8.1^{+3.1}_{-2.5}\pm 0.8)\times 10^{-6} \ \ \  \mbox{BaBar}~\cite{BrBabar}, 
\end{eqnarray}
which in fact, imply that $|A^{\mathrm{SUSY}}|/|A^{\mathrm{SM}}|$ can be of $\mathcal{O}(1)$ 
(depending on the value of the phases $\theta_{SUSY}$ and $\delta_{12}$).  
As we will show in the following sections, this is 
the right magnitude for enhancing $S_{\phi K_S}$. 

\section{{\large \bf Effective Hamiltonian for $\Delta B=1$ transitions}}
The Effective Hamiltonian for the $\Delta B=1$ transitions through penguin 
process in general can be expressed as 
\begin{equation}
\mathcal{H}^{\Delta B=1}_{\mbox{eff}}= -\frac{G_F}{\sqrt{2}}V_{tb}V_{ts}^*\left[\sum_{i=3}^{6}C_iO_i+C_gO_g
\sum_{i=3}^{6}\tilde{C}_i\tilde{O}_i+\tilde{C}_g\tilde{O}_g
\right] ,
\end{equation}
where 
\begin{eqnarray}
O_3 &=&\bar{s}_{\alpha}\gamma^{\mu}Lb_{\alpha} 
	\bar{s}_{\beta}\gamma^{\mu}Ls_{\beta}, \\ 
O_4 &=&\bar{s}_{\alpha}\gamma^{\mu}Lb_{\beta} 
	\bar{s}_{\beta}\gamma^{\mu}Ls_{\alpha}, \\ 
O_5 &=&\bar{s}_{\alpha}\gamma^{\mu}Lb_{\alpha} 
	\bar{s}_{\beta}\gamma^{\mu}Ls_{\beta}, \\ 
O_6 &=&\bar{s}_{\alpha}\gamma^{\mu}Lb_{\beta} 
	\bar{s}_{\beta}\gamma^{\mu}Rs_{\alpha}, \\
O_{g} &=& \frac{g_s}{8\pi^2}m_b\bar{s}_{\alpha}\sigma^{\mu\nu}
R\frac{\lambda^A_{\alpha\beta}}{2}b_{\beta}G^A_{\mu\nu}\label{og}.
\end{eqnarray}
where $L=1-\gamma_5$ and $R=1+\gamma_5$. 
The terms with tilde are obtained from $C_{i,g}$ and $O_{i,g}$ by exchanging $L \leftrightarrow R$.
The Wilson coefficient $C_{i(g)}$ includes both SM and SUSY contributions.  
In our analysis, we neglect the effect of the operator 
$O_{\gamma} = \frac{e}{8\pi^2}m_b\bar{s}_{\alpha}\sigma^{\mu\nu}
R b_{\alpha} F_{\mu\nu}$ and the electroweak penguin operators which 
give very small contributions.

In this paper, we follow the work by  
Ali and Greub ~\cite{AG} (the generalised factorisation approach) 
and compute the $B \to \phi K$ process by including  
the NLL order precision for the Wilson coefficients $C_{3\sim 6}$ and 
at the LL order precision for $C_g$. 
A problem of the gauge dependence and infrared singularities emerged in 
this approach was criticised in ~\cite{BurasSilves} and was discussed 
more in detail in ~\cite{ChengGF}. Improved approaches 
are available these days, i.e. the perturbative QCD approach (pQCD) 
~\cite{PQCD} and the QCD factorisation approach (BBNS) ~\cite{BBNS} and 
$B\to \phi K$ process has been calculated within those frameworks in   
~\cite{Mishima-phiK,Keum-phiK} for pQCD and ~\cite{Cheng-phiK} for BBNS. 
However, in our point of view, 
the small theoretical errors given in 
~\cite{Mishima-phiK,Keum-phiK,Cheng-phiK} are too optimistic since 
both approaches are still in progress and for instance, 
the power corrections from either $\alpha_s$ or $1/m_b$ which are not 
included in the theoretical errors of Refs. 
~\cite{Mishima-phiK,Keum-phiK,Cheng-phiK} could be sizable. 
Since the purpose of this paper is not to give any strict constraints 
on fundamental parameters of SM or SUSY, but to show if SUSY models have 
any chance to accommodate the observed deviation in CP asymmetry, 
we should stretch our theoretical uncertainties as much as possible. 
Therefore, we use Ali and Greub approach which include very large theoretical uncertainty in 
the prediction of the amplitude.  
In addition, we treat the CP conserving (strong) phase as an arbitrary value. 
Since a prediction of the strong phase is a very delicate matter, especially 
due to our ignorance of the final state interactions, 
we would like to be very conservative for its inclusion.

The Wilson coefficient at a lower scale $\mu \simeq \mathcal{O}(m_b)$  can be extrapolated by 
\begin{equation}
C_i(\mu )=\hat{U}(\mu,\mu_W)C_i(\mu_W) \ \ \ i=1\sim 6 \label{eq:0033}
\end{equation} 
where 
the evolution matrix at NLO is given by 
\begin{equation}
\hat{U}(\mu,\mu_W)=\hat{U}^{(0)}(\mu,\mu_W)+\frac{\alpha_s}{4\pi}
\left(\hat{J}\hat{U}^{(0)}(\mu,\mu_W)-\hat{U}^{(0)}(\mu,\mu_W)\hat{J}\right)
\label{NLO}
\end{equation}   
where  $\hat{U}^{(0)}$ is obtained by the 
$6\times 6$ LO anomalous dimension matrix and $\hat{J}$ is obtained by the 
NLO anomalous dimension matrix. The explicit forms of these matrices can be found 
for example, in ~\cite{Buras}. 
Since the $O_g$ contribution to $B \to \phi K_S$ is order $\alpha_s$ suppressed 
in the matrix element 
the Wilson coefficient $C_g(\mu )$  should include, for consistency, only LO corrections: 
\begin{equation}
C_g(\mu )=\hat{U}^0(\mu,\mu_W)C_g(\mu_W) \label{eq:0035}
\end{equation}
where $\hat{U}^0(\mu,\mu_W)$ is obtained by the $8\times 8$ anomalous dimension matrix of LO.

The anomalous dimension matrix at NLO  does depend on regularisation scheme. 
To avoid this 
problem, QCD corrections are carefully included in the literature ~\cite{AG}. 
As a result, the matrix element of $\overline{B}^0 \to \phi \overline{K}^0$ 
process is given by the effective Wilson coefficient, 
\begin{equation}
\langle\phi \overline{K}^0|\mathcal{H}^{\mbox{eff}}_{\Delta B=1}|\overline{B}^0\rangle
\end{equation}
where 
\begin{equation}
\mathcal{H}_{\mbox{eff}}=-\frac{G_F}{\sqrt{2}}V_{tb}V_{ts}^*
\left[\sum_{i=3}^{6} C_i^{\mbox{\small eff}}O_i
+\sum_{i=3}^{6} \tilde{C}_i^{\mbox{\small eff}}\tilde{O}_i\right]
\end{equation}
The detailed expression of the effective Wilson coefficient can be found in ~\cite{AG}. 
 The 
effective Wilson coefficient $C_i^{\mbox{eff}} (\tilde{C}_i^{\mbox{eff}})$ 
includes all the QCD corrections mentioned above. 
We must emphases that these corrections also include the contribution from the 
chromomagnetic type operator $O_g$ given in Eq. (\ref{og}). 
Note that the LLO Wilson coefficient $C_g^{\mbox{\small (SM)}}$ 
itself is an order of magnitude larger than the others, 
however it enters as a QCD corrections so that  $\alpha_s/(4\pi)\sim 1/50$ suppressed.  
As a result, the effect of $O_g$ in SM is less than 10\% level 
in the each effective Wilson coefficients  
$C_{3\sim 6}^{\mbox{\small eff(SM)}}$. 
However, we will show that in supersymmetric theories, the Wilson coefficient  
for the operator $O_g (\tilde{O}_g)$ is very large and its influence  
to the effective Wilson coefficients 
$C_{3\sim 6}^{\mbox{\small eff(SUSY)}}$  and 
$\tilde{C}_{3\sim 6}^{\mbox{\small eff(SUSY)}}$  
are quite significant. 
 
Employing the naive factorisation approximation ~\cite{AL}, where all the colour factor $N$ is 
assumed to be 3, the amplitude can be expressed as: 
\begin{equation}
\overline{A}(\phi K)=
 -\frac{G_F}{\sqrt{2}}V_{tb}V_{ts}^*\left[\sum_{i=3}^{6}C_i^{\mbox{\small eff}}
+ \sum_{i=3}^{6}\tilde{C}_i^{\mbox{\small eff}}\right]
\langle\phi\bar{K}^0 | O_i |\bar{B}^0\rangle .  
\end{equation}
The matrix element is given by: 
\begin{eqnarray}
\langle\phi\bar{K}^0 | O_3 |\bar{B}^0\rangle &=& \frac{4}{3}X, \\ 
\langle\phi\bar{K}^0 | O_4 |\bar{B}^0\rangle &=& \frac{4}{3}X, \\ 
\langle\phi\bar{K}^0 | O_5 |\bar{B}^0\rangle &=& X, \\
\langle\phi\bar{K}^0 | O_6 |\bar{B}^0\rangle &=& \frac{1}{3}X  
\end{eqnarray}
with 
\begin{equation}
X=2F_1^{B\to K}(m_{\phi}^2)f_{\phi}m_{\phi}(p_K \cdot \epsilon_{\phi}).
\end{equation}
where $F_1^{B\to K}(m_{\phi}^2)$ is the $B-K$ transition form factor and $f_{\phi}$ is the 
decay constant of $\phi$ meson. 
Note that the matrix elements for $O_{i(g)}$ and $\tilde{O}_{i(g)}$ are same for 
$B\to \phi K$ process.  
We use the following values for the parameters appearing in the above equation,
$m_{\phi} = 1.02$ GeV, $f_{\phi}=0.233$ GeV, $ (p_K \cdot \epsilon_{\phi}) = 
\frac{m_B}{m_{\phi}} \sqrt{\left[\frac{1}{2m_B}\left(m_B^2 - m_K^2 + 
m_{\phi}^2\right)\right]^2  -m_{\phi}^2} \simeq 13$ GeV, and 
$F_1^{B\to K}(m_{\phi}^2)= 0.35$ ~\cite{Ball}. 
Finally, we discuss on the matrix elements of the chromomagnetic operator $O_g$ 
which is given by: 
\begin{equation}
\langle\phi\bar{K}^0 | O_g |\bar{B}^0\rangle 
=-\frac{\alpha_s m_b}{\pi  q^2}
\left(\overline{s}_{\alpha}\gamma_{\mu}\slash{q}(1+\gamma_5)
\frac{\lambda^A_{\alpha\beta}}{2}b_{\beta}\right)
\left(\overline{s}_{\rho}\gamma^{\mu}\frac{\lambda^A_{\rho\sigma}}{2}s_{\sigma}\right)
\end{equation}
where $q^{\mu}$ is the momentum carried by the gluon in the penguin diagram. 
As discussed above, this contribution is already included in $C_{3\sim 6}^{\mbox{\small eff}}$,  
which in fact, is possible only when the matrix element of $O_g$ is written 
in terms of the matrix element of $O_{3\sim 6}$. It is achieved by using an assumption 
~\cite{AG}: 
\begin{equation}
q^{\mu}=\sqrt{\langle q^2\rangle}\frac{p_b^{\mu}}{m_b} 
\end{equation}
where $\langle q^2\rangle$ is an averaged value of $q^2$. We treat 
$\langle q^2\rangle$ as an input parameter in a range of 
$m_b^2/4< \langle q^2\rangle < m_b^2/2$. As we will see in the next section, 
our results are quite sensitive to the value of 
$\langle q^2\rangle$.

\section{{\large \bf Supersymmetric contributions to $B \to \phi K_S$ decay}}
As advocated above, the general amplitude $\overline{A}(\phi K)$  can be written as 
\begin{equation}
\overline{A}(\phi K)= \overline{A}^{\mathrm{SM}}(\phi K) +
\overline{A}^{\tilde{g}}(\phi K)+\overline{A}^{\tilde{\chi}^{\pm}}(\phi K),
\end{equation}
where $\overline{A}^{\mathrm{SM}}$, $ \overline{A}^{\tilde{g}}$, and 
$\overline{A}^{\tilde{\chi}^{\pm}}$ refer to the SM, gluino, and chargino
contributions, respectively. 
In our analysis, we consider only the 
gluino exchanges through $\Delta B=1$ penguin diagrams which give the
dominant contribution to the amplitude $\overline{A}^{\mathrm{SUSY}}(\phi K)$.
In Fig. 1, we exhibit the leading diagrams for $B \to \phi K_S$ decay.
At the first order in the mass insertion approximation, the gluino 
contributions to the Wilson coefficients $C_{i,g}$ at SUSY scale $M_S$ 
are given by
\begin{eqnarray}
C_3 (M_S) &=& \frac{\sqrt{2} \alpha_s^2}{G_F V_{tb} V_{ts}^* m_{\tilde{q}}^2} 
(\delta_{LL}^d)_{23} \left[ -\frac{1}{9} B_1(x) - \frac{5}{9} B_2(x) - 
\frac{1}{18} P_1(x) -\frac{1}{2} P_2(x) \right],\nonumber\\
C_4 (M_S) &=& \frac{\sqrt{2} \alpha_s^2}{G_F V_{tb} V_{ts}^* m_{\tilde{q}}^2}
(\delta_{LL}^d)_{23}
\left[ -\frac{7}{3} B_1(x) + \frac{1}{3} B_2(x) + \frac{1}{6} P_1(x)
+\frac{3}{2} P_2(x) \right],\nonumber\\
C_5 (M_S) &=& \frac{\sqrt{2} \alpha_s^2}{G_F V_{tb} V_{ts}^* m_{\tilde{q}}^2}
(\delta_{LL}^d)_{23}
\left[ \frac{10}{9} B_1(x) + \frac{1}{18} B_2(x) - \frac{1}{18} P_1(x)
-\frac{1}{2} P_2(x) \right], \label{eq:0047}\\
C_6 (M_S) &=& \frac{\sqrt{2} \alpha_s^2}{G_F V_{tb} V_{ts}^* m_{\tilde{q}}^2}
(\delta_{LL}^d)_{23}
\left[ -\frac{2}{3} B_1(x) + \frac{7}{6} B_2(x) + \frac{1}{6} P_1(x)
+\frac{3}{2} P_2(x) \right],\nonumber\\
C_g (M_S)\!\! &=&\!\!-\frac{\sqrt{2} \alpha_s \pi}
{G_F V_{tb} V_{ts}^* m_{\tilde{q}}^2}\!\left[
\!(\delta_{LL}^d)_{23}\left( \frac{1}{3} M_3(x)\! + \!3 M_4(x)\right)\!+\!
(\delta_{LR}^d)_{23}\frac{m_{\tilde{g}}}{m_b}
\left(\!\frac{1}{3} M_3(x) \!+\! 3 M_2(x)\right)\!\right]\!,\nonumber
\end{eqnarray}
and the coefficients $\tilde{C}_{i,g}$ are obtained from $C_{i,g}$ by 
exchanging $L \leftrightarrow R$. The functions appear in these expressions
can be found in Ref.\cite{GMS} and $x=m^2_{\tilde{g}}/m^2_{\tilde{q}}$.
As in the case of the SM, the Wilson coefficients at low energy 
$C_{i,g}(\mu)$, $\mu \simeq \mathcal{O}(m_b)$, are obtained from 
$C_{i,g}(M_S)$ by using Eqs. (\ref{eq:0033}) and (\ref{eq:0035}).

The absolute value of the mass insertions $(\delta_{AB}^d)_{23}$, with 
$A,B=(L,R)$ is constrained by the experimental results for 
the branching ratio of the  $B \to X_s \gamma$ decay \cite{GMS,bsgamma}. 
These constraints are very weak on the $LL$ and $RR$ 
mass insertions and the only limits we have come from their definition, 
$\vert (\delta_{LL,RR}^d)_{23} \vert < 1$. 
The $LR$ and $RL$ mass insertions are more constrained and 
for instance with $m_{\tilde{g}}\simeq m_{\tilde{q}}\simeq
500$ GeV, we have $\vert(\delta_{LR,RL}^d)_{23} \vert \lsim 1.6 \times 10^{-2}$. 
Although the $LR(RL)$ mass insertion is constrained so severely, 
as can be seen from the above expression of $C_{g}(M_S)$, 
it is enhanced by a large factor $m_{\tilde{g}}/m_b$. 
We will show in the following that this enhancement plays an important role 
to reproduce the observed large deviation between $\sin 2\beta$ and $S_{\phi K_S}$. 
We should recall that in the supersymmetry analysis of the direct CP violation 
in the kaon system, the same kind of enhancement by a factor $m_{\tilde{g}}/m_s$ 
makes the $LR$ and $RL$ mass insertions natural candidates to explain the experimental 
results of $\varepsilon'/\varepsilon$ \cite{KKM}. 

As shown in Eq. (\ref{eq:Sexp}), the deviation 
of $S_{\phi K_S}$ from $\sin 2\beta$ is governed by the size of  
$|A^{\mathrm{SUSY}}|/|A^{\mathrm{SM}}|$. Thus we start our analysis 
by discussing the gluino contribution to $|A^{\mathrm{SUSY}}|/|A^{\mathrm{SM}}|$. 
We choose the input parameters  as 
\begin{equation}
m_{\tilde{q}}=500\mbox{GeV}, \ x=1, \ q^2=m_b^2/2, \ \mu=2\mbox{GeV}
\end{equation}
then, we obtained 
\begin{equation}
\frac{A^{\mathrm{SUSY}}}{A^{\mathrm{SM}}} \simeq 0.13~ (\delta_{LL}^d)_{23} +
55.4~ (\delta_{LR}^d)_{23} + 55.4~ (\delta_{RL}^d)_{23} + 0.13~ 
(\delta_{RR}^d)_{23}. 
\end{equation}
The largest theoretical uncertainty comes from the choice of $q^2$. 
We find that the smaller values of $q^2$ enhance the coefficients of each mass insertions 
and for the minimum value  $q^2=m_b^2/4$ gives 
\begin{equation}
\frac{A^{\mathrm{SUSY}}}{A^{SM}} \simeq 0.23~ (\delta_{LL}^d)_{23} +
97.4~ (\delta_{LR}^d)_{23} + 97.4~ (\delta_{RL}^d)_{23} + 0.23~ 
(\delta_{RR}^d)_{23}. 
\label{Asusy}
\end{equation}
Using the constraints for each mass insertions described above, 
we obtain the maximum contribution from the individual mass insertions by 
setting the remaining three mass insertions to be zero:  
\begin{equation}
\frac{|A^{\mathrm{SUSY}}_{LL(RR)}|}{|A^{\mathrm{SM}|}}<0.23, \ \ 
\frac{|A^{\mathrm{SUSY}}_{LR(RL)}|}{|A^{\mathrm{SM}}|}<1.56. \ \ 
\end{equation} 
It is worth 
mentioning that $(\delta_{LR}^d)_{23}$ and  $(\delta_{RL}^d)_{23}$
contribute to $S_{\phi K_S}$ with the same sign, unlike their contributions
to $\varepsilon'/\varepsilon$. Therefore, in SUSY model with 
$(\delta_{LR}^d)_{23} \simeq (\delta_{RL}^d)_{23}$, we will not have the 
usual problem of the severe cancellation between their contributions, but 
we will have a constrictive interference which enhances the SUSY contribution
to $S_{\phi K_S}$. 

Now let us investigate whether 
any one of the mass insertions can accommodate  the observed  large deviation 
between $S_{J/\psi K_S}$ and $S_{\phi K_S}$.  
We start with $LL$ (same for $RR$) contribution. 
As can be seen from Eq. (\ref{eq:Sexp}), a choice of the strong phase 
$\cos\delta_{12}=\pm 1$ gives the largest deviation between $S_{\phi K_S}$ and $\sin 2\beta$. 
Using the measured central value of $\sin 2\beta$, $|(\delta_{LL}^d)_{23}|=1$ and 
$\cos\delta_{12}=\pm 1$ in the CP violation master formula Eq. (\ref{eq:Sfull}), 
we obtain $S_{\phi K_S}$ in terms of  
$\arg (\delta_{LL(RR)}^d)_{23}$ as:  
\begin{equation}
S_{\phi K_S}=\frac{0.73\pm0.45\sin (\arg(\delta_{LL(RR)}^d)_{23}+0.818)+0.051\sin (2\arg(\delta_{LL(RR)}^d)_{23}+0.818)}{1.05\pm0.45\cos(\arg(\delta_{LL(RR)}^d)_{23})}
\end{equation}
Then the minimum value of $S_{\phi K_S}$ is obtained by $\sin(\arg(\delta_{LL(RR)}^d)_{23})=\mp 0.97$ as 
\begin{equation}
S_{\phi K_S} = 0.35. 
\end{equation} 
We find that if the experimental value for $S_{\phi K_S}$ remains as small as the 
current central value, the SUSY models with $LL$ or $RR$ mass insertion can not provide 
an explanation for that. 

Next, we show that on the contrary, the $(\delta_{LR(RL)}^d)_{23}$ contribution can deviate 
$S_{\phi K_S}$ from $\sin 2\beta$ much more significantly. 
For $|(\delta_{LR(RL)}^d)_{23}|=0.01$ and 
$\cos\delta_{12}=\pm 1$, the mixing CP asymmetry is expressed as  
\begin{equation}
S_{\phi K_S}=\frac{0.73\pm1.95\sin (\arg(\delta_{LR(RL)}^d)_{23}+0.818)+0.95\sin (2\arg(\delta_{LR(RL)}^d)_{23}+0.818)}{1.95\pm1.95\cos(\arg(\delta_{LR(RL)}^d)_{23})}. 
\end{equation}
and the minimum value is obtained with $\sin(\arg(\delta_{LR(RL)}^d)_{23})=(\mp 0.075,\  \mp 0.63)$ as 
\begin{equation}
S_{\phi K_S} = -1. 
\end{equation} 

\begin{figure}[t]
\psfrag{(a)}[r][r][0.3]{(a)}\psfrag{(c)}[r][r][0.3]{(b)}
\psfrag{s}[r][l][0.3]{$S_{\phi K_S}$}
\psfrag{1ll}[l][l][0.2]{\large $|(\delta_{LL(RR)}^d)_{23}|=0.1$}
\psfrag{2ll}[l][l][0.2]{\large $|(\delta_{LL(RR)}^d)_{23}|=0.5$}
\psfrag{3ll}[l][l][0.2]{\large $|(\delta_{LL(RR)}^d)_{23}|=1$}
\psfrag{1lr}[l][l][0.2]{\large $|(\delta_{LR(RL)}^d)_{23}|=0.001$}
\psfrag{2lr}[l][l][0.2]{\large $|(\delta_{LR(RL)}^d)_{23}|=0.005$}
\psfrag{3lr}[l][l][0.2]{\large $|(\delta_{LR(RL)}^d)_{23}|=0.01$}
\psfrag{ll}[l][l][0.35]{$\arg[(\delta_{LL(RR)}^d)_{23}]$}
\psfrag{lr}[l][l][0.35]{$\arg[(\delta_{LR(RL)}^d)_{23}]$}
\begin{center} \scalebox{1.6}{
\includegraphics[width=10cm,height=5cm,keepaspectratio]{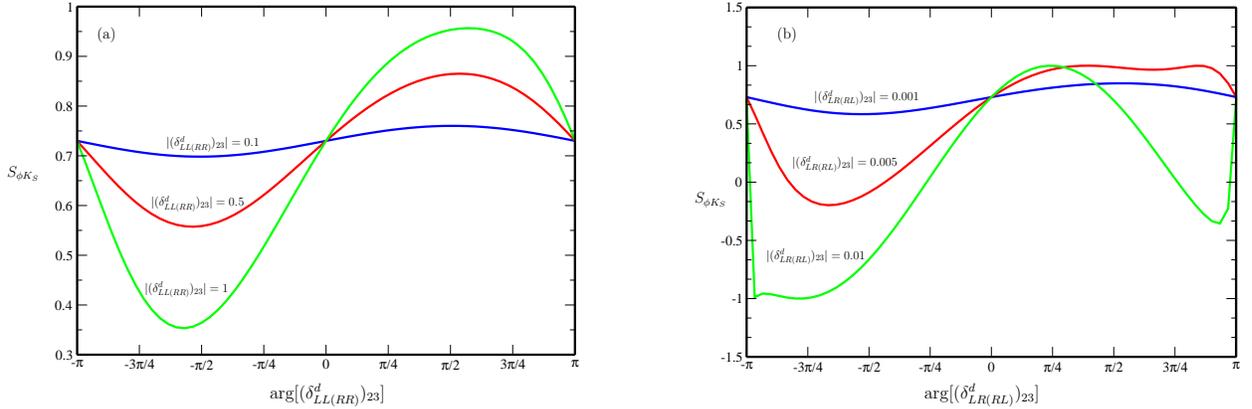}}
\caption{{\small The mixing CP asymmetry as function of $\arg[(\delta_{AB}^d)_{23}]$ for 
three values of the $|(\delta_{AB}^d)_{23}|$ where $AB=LL(RR) \mbox{(a)},\   
LR(RL) \mbox{(b)}$. The strong phase $\delta_{12}$ is fixed at $\cos\delta_{12}=1$. }}
\label{Fig:2}
\end{center}
\end{figure}

In Fig.{\ref{Fig:2}}, we present plots for 
the phase of $(\delta_{LL(RR)}^d)_{23}$ and $(\delta_{LR(RL)}^d)_{23}$ versus
the mixing CP asymmetry $S_{\phi K_S}$ for  
$\cos\delta_{12}=1$. 
We choose the three values of the magnitude of these mass 
insertions within the 
bounds from the experimental limits in particular, from  $B \to X_s \gamma$.  
Each plot shows a contribution from an individual mass insertion by setting the 
other three to be zero. 
As can be seen from these plots, the $LR$ (same for $RL$) gives the largest contribution 
to $S_{\phi K_S}$.  In order to have 
a sizable effect from the $LL$ or $RR$, 
the magnitude of  $(\delta_{LL(RR)}^d)_{23}$ has to be of order one 
and furthermore, the imaginary part needs to be as large as the real part.   
In any case, it is very difficult to give negative value of $S_{\phi K_S}$ from $(\delta_{LL}^d)_{23}$ or $(\delta_{RR}^d)_{23}$ mass insertion.  
On the contrary, even if we reduce the magnitude of 
$(\delta_{LR}^d)_{23}$ to the half of its maximum value, $S_{\phi K_S}$ can still reach to a negative value. 
We also find that in the case of $|(\delta^d_{LR(RL)})_{23}|=0.01$, 
the minimum value of $S_{\phi K_S}$ can be achieved without large imaginary part. 
Actually, we should notify that smaller values of the ratio 
$x=m_{\tilde{g}}^2/m_{\tilde{q}}^2$ can lead to 
large $LL$ and $RR$ contributions to $S_{\phi K_S}$. However, 
the negative value of $S_{\phi K_S}$ can be only achieved by  
the light gluino mass around the current experimental limit and 
very heavy (order of TeV) squark mass.   

\begin{figure}[h]
\psfrag{s}[r][l][1]{$S_{\phi K_S}$}\psfrag{c}[r][l][1]{$C_{\phi K_S}$}
\psfrag{miss}[r][l][0.7]{$\pi/2$}
\begin{center} 
\includegraphics[width=10cm,height=10cm,keepaspectratio]{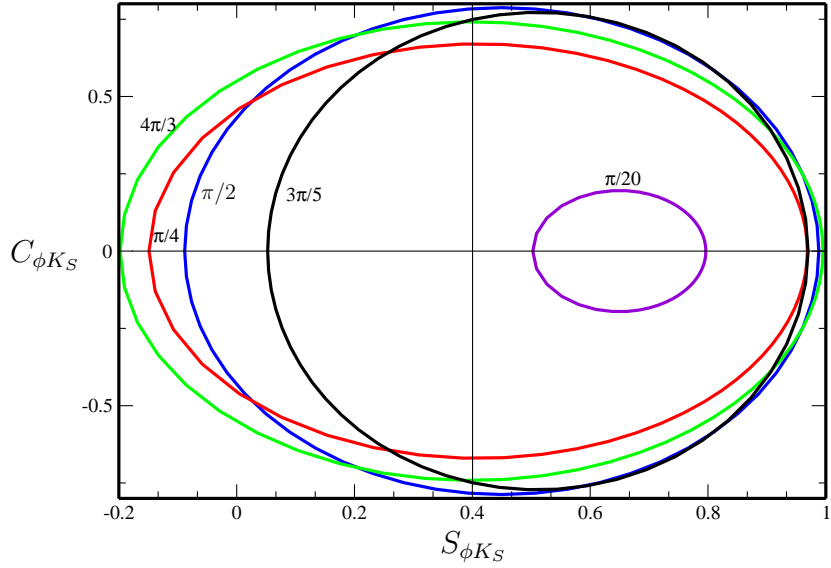}
\caption{{\small The mixing CP asymmetry $S_{\phi K_S }$ versus 
the direct CP symmetry $C_{\phi K_S}$ for strong phase $\delta_{12} \in [0,2\pi]$ and 
five representative values of $\theta_{SUSY}=(\pi/20, \pi/4, \pi/2, 3\pi/5, 4\pi/3)$.}}
\label{figure3}
\end{center}
\end{figure}

So far, we have only considered the cases where the strong phase $\delta_{12}$ is given by 
$\cos\delta_{12}=\pm 1$ 
so that the direct CP asymmetry $C_{\phi K_S}$ was identically zero.
However, since the correlation between $S_{\phi K_S}$ and $C_{\phi K_S}$ would 
be very useful to give  some constrains on SUSY parameters, 
here we include arbitrary strong phases and demonstrate an example of 
$S_{\phi K_S}-C_{\phi K_S}$ correlation.  
By using the expanded formulae for $S_{\phi K_S}$ and $C_{\phi K_S}$ in Eqs. (\ref{eq:Sexp})
and (\ref{eq:Cexp}), we  find that for any value of $\delta_{12}$,  
the plot of $S_{\phi K_S}$ versus $C_{\phi K_S}$ becomes an ellipse with its size proportional to 
$\sin\theta_{SUSY}$:  
\begin{equation}
\frac{(S_{\phi K_S}-\sin 2\beta)^2}{\cos^22\beta\left(2\sin\theta_{SUSY}
\frac{|A^{\mathrm{SUSY}}|}{|A^{\mathrm{SM}}|}\right)^2}+
\frac{C_{\phi K_S}^2}
{\left(2\sin\theta_{SUSY}\frac{|A^{\mathrm{SUSY}}|}{|A^{\mathrm{SM}}|}\right)^2}=1
\end{equation}  
In Fig.\ref{figure3}, we depict an example of the plot with $|A^{\mathrm{SUSY}}|/|A^{\mathrm{SM}}|\simeq 0.5$ 
and $\theta_{SUSY}=(\pi/20, \pi/4, \pi/2, 3\pi/5, 4\pi/3)$. Since we used the 
full formulae in Eqs. (\ref{eq:Sfull}) and (\ref{eq:Cfull}) to create this figure, 
different $\theta_{SUSY}$ does not give precisely rescaled ellipses. 
Nevertheless we can see the qualitative feature. 
The strong phase $\delta_{12}=0$ corresponds to the point at the right most tip of the ellipse. 
As $\delta_{12}$ increases, it runs anti-clockwise and finishes a round when 
$\delta_{12}=2\pi$. 
Note that with $|A^{\mathrm{SUSY}}|/|A^{\mathrm{SM}}|\simeq 0.5$, 
a typical branching ratio is about $10\times 10^{-6}$, which is within the 
range of experimental value.  
As experimental errors  will be reduced, this kind of plot would provide 
very interesting constraints on SUSY parameters. 

\vspace*{0.8cm}

\section{{\large \bf CP asymmetry $S_{\phi K_S}$ in explicit SUSY models}}

In this section, we study the CP asymmetry of the $B\to \phi K_S$ in 
some specific SUSY models. 
We consider the minimal supersymmetric standard model (MSSM) 
(where minimal number of superfields is introduced and $R$ parity is 
conserved) with the following soft SUSY breaking terms
\bea\label{susy:gen:vsb}
V_{SB} &=& m_{0\alpha}^2 \phi_{\alpha}^* \phi_{\alpha} +
\epsilon_{ab}
(A^u_{ij} Y^u_{ij} H_2^b \tilde{q}_{L_i}^a \tilde{u}^*_{R_j} +
A^d_{ij} Y^d_{ij} H_1^a \tilde{q}_{L_i}^b \tilde{d}^*_{R_j} +
A^l_{ij} Y^l_{ij} H_1^a \tilde{l}_{L_i}^b \tilde{e}^*_{R_j} \nonumber\\
&-& B\mu H_1^a H_2^b + \mathrm{H.c.})
- \frac{1}{2}
(m_3\bar{\tilde{g}} \tilde{g} +
m_2 \overline{\widetilde{W^a}} \widetilde{W}^a +
m_1 \bar{\tilde{B}} \tilde{B})\;,
\eea
where $i,j$ are family indices, $a,b$ are $SU(2)$ indices, 
$\epsilon_{ab}$ is the $2\times 2$ fully antisymmetric tensor, with
$\epsilon_{12}=1$, and $\phi_{\alpha}$ denotes all
the scalar fields of the theory.
We start with minimal supergravity model and then we consider 
general SUSY models with non--universal soft breaking terms. We also 
discuss the impact of the type of Yukawa couplings on the prediction of the 
later model.

\subsection{{\large \bf minimal supergravity model}}
In a minimal supergravity framework, the soft SUSY breaking
parameters are universal at GUT scale, and we can write
\begin{equation}\label{susy:gen:msugra}
m_{0\alpha}^2 = m_0^2 \;, \quad m_i=m_{1/2}\;, \quad
A^\alpha_{ij} = A_0 e^{i \phi_A}\;.
\end{equation}
In this model, there are only two physical phases: $\phi_{A} = \arg(A^* m_{1/2})$ and 
$\phi_{\mu}=\arg(\mu m_{1/2})$. In order to have EDM values below the experimental 
bounds, and without forcing the SUSY masses to be unnaturally heavy,
the phases $\phi_{A}$ and $\phi_{\mu}$ must be at most of order 
$10^{-1}$ and $10^{-2}$, respectively~\cite{phase:edm}.

It is clear that this class of models, where the SUSY phases are constrained
to be very small and the Yukawa couplings are the main source of the flavour
structure, can not generate sizable contributions to the CP violating 
parameters. And it is indeed the case for $S_{\phi K_S}$. 
We find that it is impossible to have a large deviation between 
$S_{\phi K_S}$ and $\sin 2\beta$ within the minimal supergravity framework. 

In fact, we find that even if we ignore the bounds from the EDM, and 
allow large values for SUSY phases, $\phi_{A,\mu} \simeq \pi/2$, still 
the SUSY contribution to  $S_{\phi K_S}$ is negligible. The suppression
is mainly due to the universality assumption of the soft breaking terms. 
For instance, with $m_{1/2}\simeq m_0\simeq A_0 \simeq 200$ GeV we find 
the following values of the relevant mass insertions:
\bea
(\delta^d_{LL})_{23} &\simeq& 0.009 + i~ 0.001,\\
(\delta^d_{RR})_{23} &\simeq& - 2.1 \times 10^{-7} - i~ 2.5 \times 10^{-8},\\
(\delta^d_{LR})_{23} &\simeq& - 2.5\times 10^{-5} - i~ 1.9 \times 10^{-5}.
\eea
Clearly these values are much smaller than the corresponding values 
mentioned in the previous section and it gives negligible 
contributions to the CP asymmetry $S_{\phi K_S}$.
Indeed, we find that the total $S_{\phi K_S}$ in this example is given by 
$S_{\phi K_S} =0.729 $, which is about the value of $\sin 2\beta$.

\subsection{SUSY models with non--universal soft terms}
Now we consider SUSY models with non--universal soft terms. In particular, we 
focus on the models with non--universal $A$--terms in order to enhance the 
values of $(\delta_{LR}^d)_{23}$ and $(\delta_{RL}^d)_{23}$, which may give 
the dominant contributions to the CP asymmetry $S_{\phi K_S}$. However, 
non--observation of EDMs leads to restrictive constraints on the 
non--degenerate $A$--terms and only certain patterns of flavour 
are allowed, such as the Yukawa and $A$--terms are Hermitian \cite{hermitian},
or the $A$--terms are factorisable, {\it i.e.}, $(Y^A)_{ij} = A.Y$ or $Y.A$
\cite{KKM}. In the case of factorisation, the mass insertion 
$(\delta^d_{LR})_{11}$ is suppressed by the ratio $m_d/m_{\tilde{q}}$. Here 
we will consider this scenario with the following trilinear structure:
 \begin{equation}\label{Aterms}
A = m_0 \left(\begin{array}{ccc}
a & a & a \\
b & b & b \\
c & c & c \end{array} \right) \;.
\end{equation}
As pointed out in Ref.\cite{ABK}, in the case of non--universal soft
breaking terms, the type of the Yukawa couplings (hierarchical or nearly 
democratic) plays an important role and has significant impact on the 
predictions of these models. If we consider the standard example of  
hierarchical quark Yukawa matrices,
\begin{eqnarray}
Y_{u} & = & \frac{1}{v\sin \beta }\mathrm{diag}\left( m_{u},m_{c},
m_{t}\right) ,\nonumber \\
Y_{d} & = & \frac{1}{v\cos \beta }K.\mathrm{diag}\left( m_{u},m_{c},
m_{t}\right) .K^{+},
\end{eqnarray}
 where \( K \) is the CKM matrix, the relevant $LR$ mass insertion is 
given by
\be
(\delta_{LR}^d)_{23} \simeq \frac{v\cos\beta}{m_{\tilde{q}^2}} \left(K^+~ Y^A_d~ K
\right)_{23} 
\ee
where $(Y_d^A)_{ij}=Y^d_{ij} A^d_{ij}$. It is clear that the dominant 
contribution to this mass insertion is given by the term 
$K_{22} (Y_d^A)_{23} K_{33}^+$ which is still suppressed by the small entry of 
$Y^d_{23}$. The non--universality of the squarks can enhance the $LL$ and $RR$
mass insertions, however this non--universality is severely constrained 
by the experimental measurements of $\Delta M_{K}$ and $\varepsilon_{K}$.
Therefore, with the hierarchical Yukawa couplings  we find that the 
typical values of the relevant mass insertions
are at least two order of magnitude below the required values so that 
contributions to split the CP asymmetries $S_{J/\psi K_S}$ from $S_{\phi K_S}$
are again small.

Now we consider the same SUSY model but with converting the above 
hierarchical Yukawa matrices to democratic ones, which can be obtained
by a unitary transformation. As emphasised in Ref.\cite{ABK} that these
new Yukawa textures (and their diagonalising matrices $S^{u,d}_{L,R}$)
have large mixing, which has important consequences in the SUSY results. 
Thus, the element $(Y_d^A)_{23}$ has no suppression factor as before and 
the magnitude of $(\delta_{LR}^d)_{23}$ can be of the desired order. 
As a numerical example, for $m_{0}=m_{1/2}=200$ GeV, 
({\it i.e.}, $m_{\tilde{q}} \simeq m_g \simeq 500$ GeV)
and assuming that $\vert A_{ij} \vert \in [m_0, 4 m_0]$ while 
the phases of the $A$--terms are chosen such that the 
bound of the EDMs are satisfied, 
we find that it is quite natural to obtain the following values of the 
mass insertion $(\delta_{LR}^d)_{23}$: 
$\vert (\delta_{LR}^d)_{23} \vert \simeq 0.005$ and
$\mathrm{Arg}[(\delta_{LR}^d)_{23}] \simeq 1.2$ which leads to $S_{\phi K_S}
\simeq - 0.2$. 

\section{Conclusions}
In this paper, we have studied the supersymmetric contributions to the CP
asymmetry of $B \to \phi K_S$ process. Using the mass insertion
approximation method, we have derived model independent limits for the
mixing CP asymmetry $S_{\phi K_S}$. We found that the $LR$ or $RL$ mass insertion
gives the largest contribution to $S_{\phi K_S}$, while the $LL$ or $RR$
contribution is small. Thus, we conclude that 
if the deviation between $S_{\phi K_S}$ and $S_{J/\psi K_S}$ observed by
the B--factory experiments (Belle and BaBar) remains as large as its
present central value, the SUSY models with large ($\sim 10^{-3}$) $LR(RL)$ mass
insertions will be an interesting candidate to explain this phenomena.  

The Belle collaboration observed non--vanishing direct CP asymmetry
$C_{\phi K_S}$ which can be obtained only by simultaneous non--vanishing
strong phase and SUSY CP violating phase. Thus, we studied the impact of the
strong phase in our results for $S_{\phi K_S}$. We provided an example of 
a plot to show a correlation between $S_{\phi K_S}$ and $C_{\phi K_S}$. 
Our result will be useful to give constraints on some SUSY parameters 
as experimental errors will be reduced. 

We also applied our results to the minimal supergravity model and SUSY
models with non--universal soft terms with two types of Yukawa couplings,
hierarchal and nearly democratic Yukawa textures. We showed that
only in SUSY models with large Yukawa mixing, SUSY contributions 
could be enhanced and reach the desired values to give
significant contributions to the CP asymmetry $S_{\phi K_S}$. This result
motivates the interest in SUSY models with non--universal soft terms and
also shed the light on the type of the Yukawa flavour structure.

\section*{\bf \normalsize Note added in proof} 
We have found an error in our program and the $RR$ contribution is 
enhanced by one order of magnitude in this revised version. 
As a result, we agree with the results in Refs. ~\cite{Murayama,Masiero}. 

\section*{\bf \normalsize Acknowledgements}
We would like to thank Patricia Ball and Jean-Marie Fr\`ere for useful 
discussions. Communication with Hitoshi Murayama 
which lead us to find an error in our program is gratefully acknowledged.  
This work was supported by PPARC. 

%

\end{document}